\pdfoutput=1
\documentclass[amsmath,prd,aps,twocolumn,showpacs,superscriptaddress,floatfix,nofootinbib]{revtex4}

\usepackage{amsfonts}
\usepackage{bm}
\usepackage{color}
\usepackage{graphicx}
\definecolor{darkgreen}{cmyk}{0.85,0.2,1.00,0.2}

\newcommand{\be}{\begin{equation}}
\newcommand{\ee}{\end{equation}}
\newcommand{\ba}{\begin{eqnarray}}
\newcommand{\ea}{\end{eqnarray}}

\newcommand\lsim{\mathrel{\rlap{\lower4pt\hbox{\hskip1pt$\sim$}}
        \raise1pt\hbox{$<$}}}
\newcommand\gsim{\mathrel{\rlap{\lower4pt\hbox{\hskip1pt$\sim$}}
        \raise1pt\hbox{$>$}}}

\def\kpiv{k_0}
\def\ellmin{\ell_{\rm min}}
\def\ellmax{\ell_{\rm max}}
\def\fsky{f_{\rm sky}}

\begin{document}

\title{On testing and extending the inflationary consistency relation for tensor modes}

\author{Latham Boyle}
\affiliation{Perimeter Institute for Theoretical Physics, Waterloo ON N2L 2Y5}
\author{Kendrick M.~Smith}
\affiliation{Perimeter Institute for Theoretical Physics, Waterloo ON N2L 2Y5}
\author{Cora Dvorkin}
\affiliation{Institute for Advanced Study, School of Natural Sciences, Einstein Drive, Princeton, NJ 08540, USA}
\author{Neil Turok}
\affiliation{Perimeter Institute for Theoretical Physics, Waterloo ON N2L 2Y5}

\date{\today}


\begin{abstract}
If observations confirm BICEP2's claim of a tensor-scalar ratio $r\approx 0.2$ on CMB scales, then the inflationary consistency relation $n_{t}=-r/8$  predicts a small negative value for the tensor spectral index $n_t$. We show that future CMB polarization experiments should be able to confirm this prediction at several sigma. We also show how to properly extend the consistency relation to solar system scales, where the primordial gravitational wave density $\Omega_{gw}$ could be measured by proposed experiments such as the Big Bang Observer. This would provide a far more  stringent test of the consistency relation and access much more detailed information about the early universe.
\end{abstract}


\maketitle


The BICEP2 experiment has recently claimed a detection of $B$-mode
cosmic microwave background (CMB) polarization on large angular 
scales~\cite{Ade:2014xna}. The observed peak in $B$-mode power at $\ell\approx 60$ is consistent
with the predicted signal from a background of gravitational waves generated
quantum mechanically during inflation, with tensor-to-scalar ratio $r \approx 0.2$.
For the simplest models of inflation, namely single-field models satisfying
the slow-roll conditions, this value of $r$ corresponds to inflationary
energy scale $V^{1/4}\approx 2.2 \times 10^{16}$ GeV and an inflaton field 
excursion $\Delta\phi \approx 10 M_{\rm Pl}\approx 2.4 \times 10^{19}$ GeV.

Subsequent analyses \cite{Mortonson:2014bja, Flauger:2014qra} (including, in particular, 
the BICEP2$\times$Planck cross correlation analysis \cite{Ade:2015tva}) have found that
at least a significant fraction of the BICEP2 B-mode signal is likely due to dust.  
Nevertheless, this signal may still contain a significant primordial inflationary contribution; 
and it is natural to explore further tests of this possibility, and ask what future measurements 
might be made to characterize or constrain gravitational waves from the very early universe.  

On CMB scales, the tensor power spectrum is predicted to be almost a
power-law, $P_t(k) \propto k^{n_t}$, in nearly all models of inflation.
In single-field slow-roll models, the spectral index $n_t$ satisfies the
``consistency relation'':
\be
n_t = -r/8.  
\label{eq:consistency}
\ee
Therefore, if BICEP2 has indeed detected primordial tensor modes, a natural target 
for future CMB experiments is a tensor tilt of order $n_t\approx -0.025$. Unfortunately, 
the sample variance limit for an all-sky ideal $B$-mode measurement is $\sigma(n_t) 
\approx 0.03$. This limit comes from the sample variance of both the gravitationally 
lensed $B$-modes  on scales $\ell \gsim 150$, and the primordial gravity wave 
$B$-modes for  $\ell \lsim 150$. The first of these contributions can be reduced 
through ``delensing'' algorithms which statistically separate lensed and gravity
wave $B$-modes.  In the first section of this paper, we will explore prospects for 
using delensing to verify the single-field slow roll consistency relation, assuming that 
the gravity wave amplitude is as large as BICEP2 suggests.

The consistency relation~(\ref{eq:consistency}) is a prediction for the scale
dependence of the gravity wave amplitude over the range of scales observable in the 
CMB (roughly $10^{24}$--$10^{27}$ m). There is a second window of scales where 
we might observe cosmological gravity waves: on solar system scales of order $10^{9}$ m, 
using interferometers such as the proposed Big Bang Observer (BBO).  Together, these 
measurements of $P_{t}(k)$ span a factor of $10^{18}$ (more than 40 efolds) in scale,
providing a huge lever arm to test the slight scale dependence predicted by 
inflation.    However, the prediction~(\ref{eq:consistency}) must be extended, 
since the power-law form $P_t(k) \propto k^{n_t}$ is no longer a good approximation over 
such a vast range.  In the second section of this paper, we show how to reformulate 
Eq.~(\ref{eq:consistency}) as a single-field consistency relation which extends all the way 
down to solar system scales, and we explore the prospects for testing it with future 
space-based interferometric experiments.

The earlier papers \cite{Caligiuri:2014sla,Dodelson:2014exa} have also considered related
aspects of delensing and direct detection of the primordial tensor spectrum in light of 
the BICEP2 results.

\section{CMB delensing}

\begin{figure}
\centerline{\includegraphics[width=6.74cm]{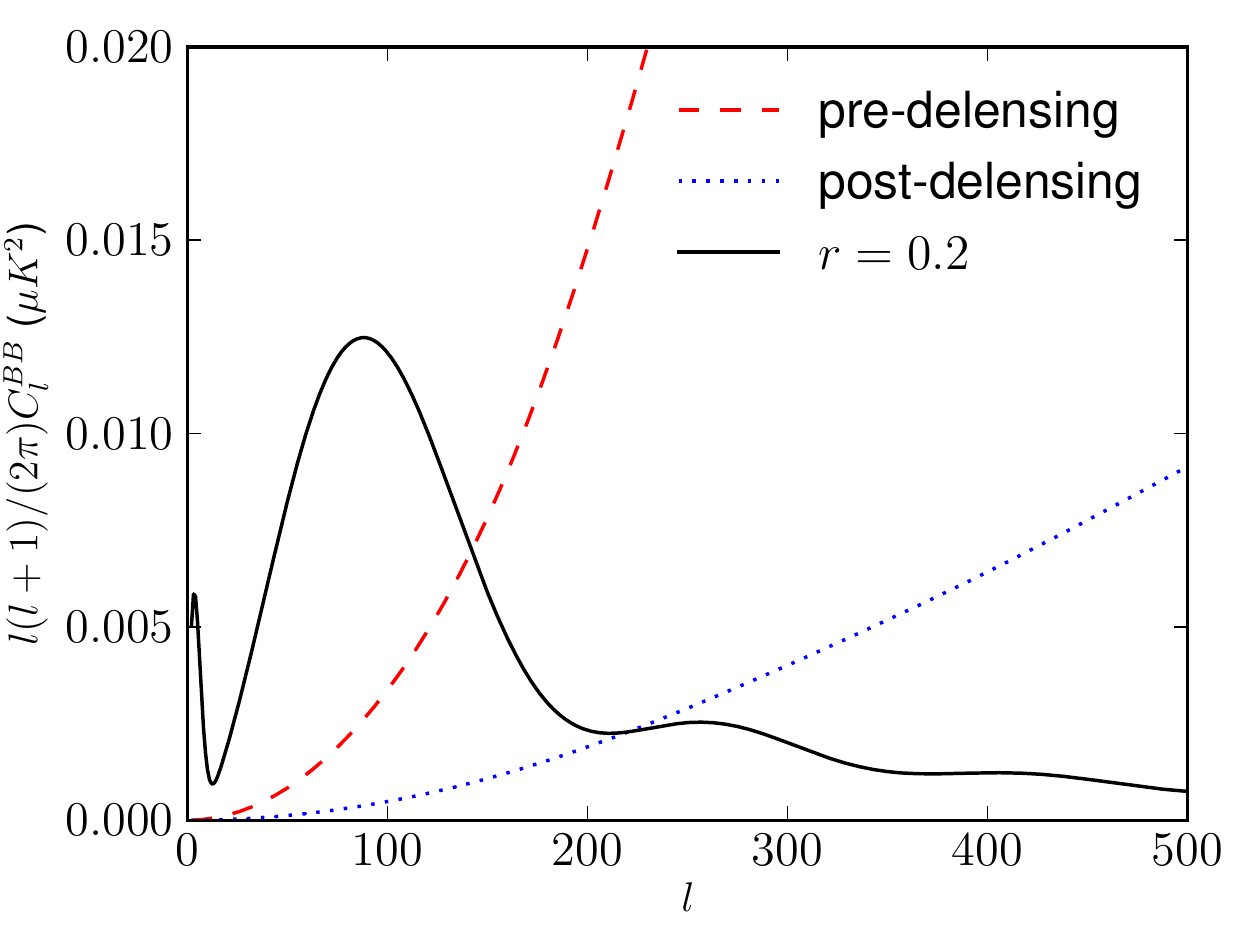}}
\caption{Gravity wave and lensing contributions to $C_\ell^{BB}$,
before and after delensing with $\Delta_P = 1$ $\mu$K-arcmin and 
$\theta_{\rm FWHM} = 2$ arcmin. Delensing allows the gravity wave signal
to be measured to higher $\ell$, improving the lever arm for measuring $n_t$.}
\label{fig:delensed_bb}
\end{figure}

The intuitive idea of delensing is to use higher-point correlations to statistically
separate lensed $B$-modes (which are non-Gaussian) and Gaussian gravity wave $B$-modes.
Delensing mixes scales in such a way that measurements of {\em small-scale}
$E$-modes and $B$-modes are used to delense large-scale $B$-modes.
For a detailed description, see~\cite{Knox:2002pe,Kesden:2002ku,Seljak:2003pn}.
For forecasting purposes, we can simply consider delensing to be a procedure
which reduces the effective lensing contribution to $C_\ell^{BB}$.
This is illustrated in Fig.~\ref{fig:delensed_bb}, where the
lensed B-mode power spectrum is compared to the power spectrum of the residual
lensed B-modes after delensing, in an experiment with noise level
$\Delta_P=1$ $\mu$K-arcmin and beam $\theta_{\rm FWHM} = 2$ arcmin.
All delensing results in this paper use the forecasting methodology from~\cite{Smith:2010gu}
(with $\ellmax=4000$).

\begin{figure}
\centerline{\includegraphics[width=6.5cm]{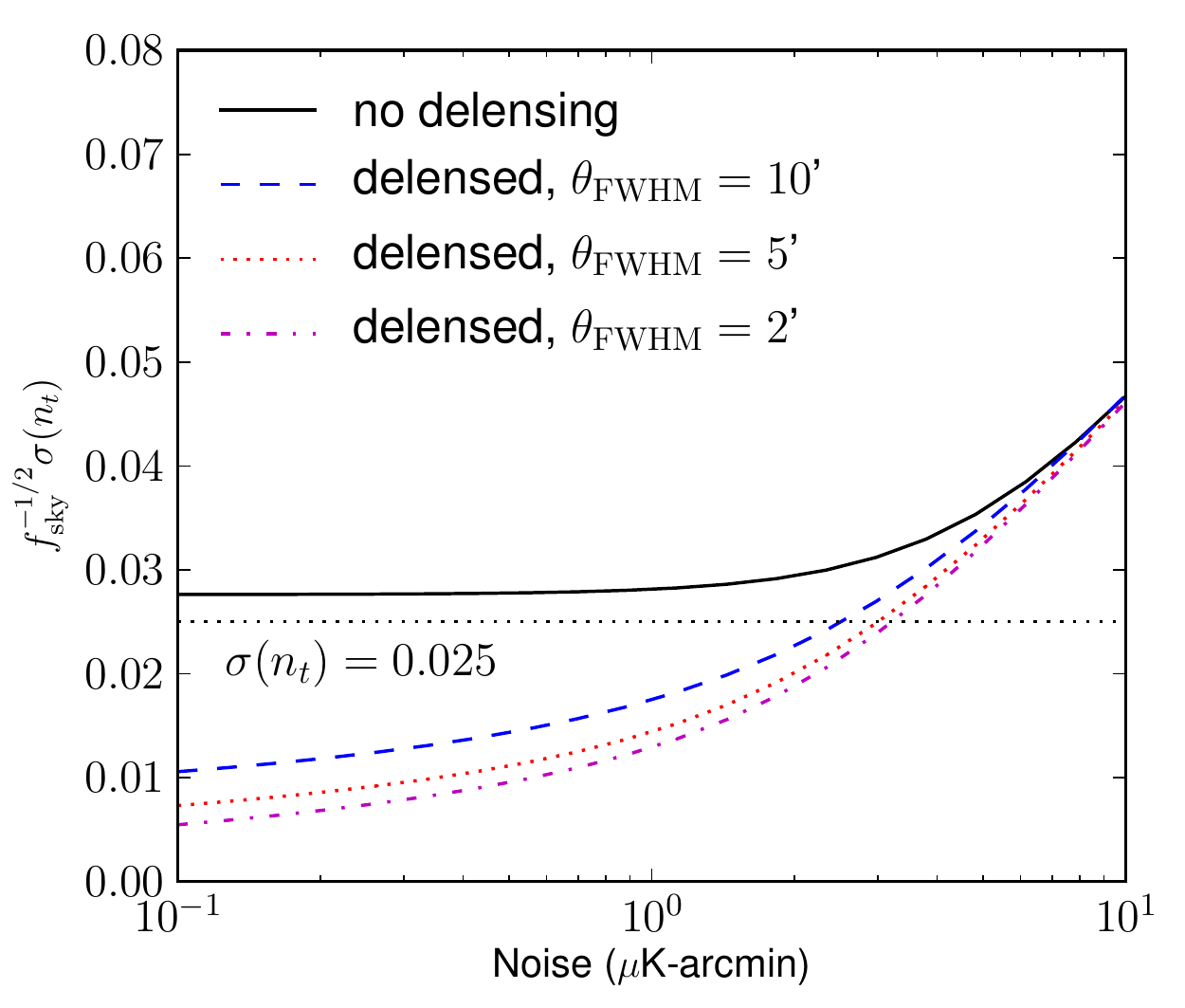}}
\caption{Forecasted statistical error $\sigma(n_t)$, marginalized over $r$,
for varying instrumental noise $\Delta_P$, with or without 
delensing, computed using the Fisher matrix 
and noise power spectrum estimation machinery from~\cite{Smith:2010gu}.}
\label{fig:delensing_forecasts}
\end{figure}

In an $r=0.2$ world, delensing is no longer interesting as a way of reducing statistical errors on $r$,
but potentially very interesting as a way of reducing statistical errors on $n_t$.
In Fig.~\ref{fig:delensing_forecasts}, we forecast the statistical error $\sigma(n_t)$ marginalized over $r$,
for varying noise and beam, with and without delensing.
A very ambitious experiment with $\Delta_p\lsim1$ $\mu$K-arcmin, a few-arcminute beam, and $\fsky\sim1$  could test the consistency relation at a few sigma. If ground-based, the number of detectors required for such an experiment is 
$\sim10^6$~\cite{Wu:2014hta}. Experiments with roughly these specifications have already been proposed.

An interesting aspect of the above forecast is that at noise levels where delensing helps,
the delensed statistical error on $n_t$ is nearly unchanged if we take $\ellmin=25$, rather than $\ellmin=2$ as has been assumed in Fig.~\ref{fig:delensing_forecasts}. Thus, for constraining $n_t$, it is not necessary to measure the reionization bump at $\ell \lsim 10$. It is also interesting to ask, how should $\fsky$ be chosen in order to minimize the statistical error $\sigma(n_t)$, assuming that the total sensitivity $(\Delta_p \fsky^{-1/2})$ is held fixed? For surveys whose total sensitivity is at least as good as BICEP2, we find that $\sigma(n_t)$ is always a decreasing function of $\fsky$. Taken together, these observations suggest that the optimal strategy for constraining $n_t$ is to observe many scattered patches of a few hundred square degrees or larger, with sky locations chosen to minimize astrophysical foregrounds.

The tensor-to-scalar ratio $r$ is defined for a given ``pivot'' wavenumber $\kpiv$ as
$r = P_t(\kpiv) / P_s(\kpiv)$, so it depends on pivot wavenumber as $r \propto \kpiv^{n_t-(n_s-1)}$.
We briefly discuss the dependence of the above forecasts on the choice of $\kpiv$.
The correlation coefficient $\rho = \mbox{Corr}(r,n_t)$ depends on the choice of $\kpiv$, but the $r$-marginalized error $\sigma(n_t)$ does not. If $\kpiv$ is chosen close to the scale where $n_t$ is best measured, as appropriate for verifying the consistency relation~(\ref{eq:consistency}), then $\rho$ will be small and the statistical error on $n_t$ will be the same regardless of whether $r$ is marginalized.
Therefore, for purposes of verifying the consistency relation, the $r$-marginalized error $\sigma(n_t)$ is 
always the correct quantity to consider, but the tilt of the error ellipse in the $(r,n_t)$ plane depends on the choice of pivot wavenumber.

\section{Direct GW detection}

If the primordial gravitational wave background is as large as BICEP2 suggests, it can also be 
directly detected by a space-based gravitational wave detector like the proposed Big Bang 
Observer (BBO) mission \cite{BBOproposal, Cutler:2005qq, Corbin:2005ny}, or perhaps even the 
somewhat less sensitive 
DECIGO mission \cite{Kawamura:2006up, Kawamura:2011zz}.  For other work relating CMB
and BBO constraints, see \cite{Turner:1996ck, Ungarelli:2005qb, Kuroyanagi:2009br, 
Kuroyanagi:2011iw, Smith:2014kka, Jinno:2014qka, Smith:2005mm, Boyle:2005se, Smith:2006xf, 
Boyle:2007zx, Chongchitnan:2006pe, Caligiuri:2014ola}. These two laser interferometer (LI) missions are designed to detect gravitational waves of present-day frequency $f\sim0.3~{\rm Hz}$, 
since this is (roughly) the lowest frequency that is uncontaminated by the gravitational 
wave foreground from white dwarf binaries. 

To translate CMB observations into a prediction for $\Omega_{gw}$ (the present-day strength of the 
relic gravitational wave background on LI scales), one proceeds in two steps: (i) first, one
extrapolates the primordial tensor power spectrum $P_{t}(k)$ from CMB to LI scales;
and (ii) second, one propagates these primordial tensors to the present time with the 
tensor transfer function.  Let us consider these two steps in turn.

{\bf Step (i): Extrapolating from CMB to LI scales:}  The traditional approach \cite{Turner:1996ck, 
Ungarelli:2005qb, Kuroyanagi:2009br, Kuroyanagi:2011iw, Smith:2014kka, Jinno:2014qka} is 
to expand ${\rm ln}(P_{t})$ as a Taylor series in ${\rm ln}(k/k_{0})$ around the wavenumber $k_{0}$:
\begin{equation}
  \label{lnkExpansion}
  {\rm ln}\frac{P_{t}(k)}{P_{t,0}}=\frac{n_{t}}{1!}({\rm ln}\frac{k}{k_{0}})^{1}+\frac{\alpha_{t}}{2!}
  ({\rm ln}\frac{k}{k_{0}})^{2}+\frac{\beta_{t}}{3!}({\rm ln}\frac{k}{k_{0}})^{3}+\ldots
\end{equation}
and then use the inflationary consistency relations to re-express the first few Taylor coefficients in terms of 
the CMB observables $\tilde{r}\equiv r/8$, $\delta n_{s}\equiv n_{s}-1$ and $\alpha_{s}$ as follows:
$n_{t}\approx-\tilde{r}$, $\alpha_{t}\approx\tilde{r}(\delta n_{s}+\tilde{r})$, $\beta_{t}\approx
\tilde{r}(\alpha_{s}-\delta n_{s}^{2}-3\tilde{r}\delta n_{s}-2\tilde{r}^{2})$.  These expressions are
valid to leading order in slow roll: we have carefully checked that the next-to-leading order (NLO)
corrections are unnecessary in the traditional approach, as they yield a negligible improvement 
in the extrapolation accuracy.  (For help in deriving these expressions to leading order or NLO, 
see Refs.~\cite{Lidsey:1995np, Cortes:2006ap}.)  

\begin{figure}
\centerline{\includegraphics[width=7.5cm,clip=True,trim=0.5cm 4cm 0.5cm 4cm]{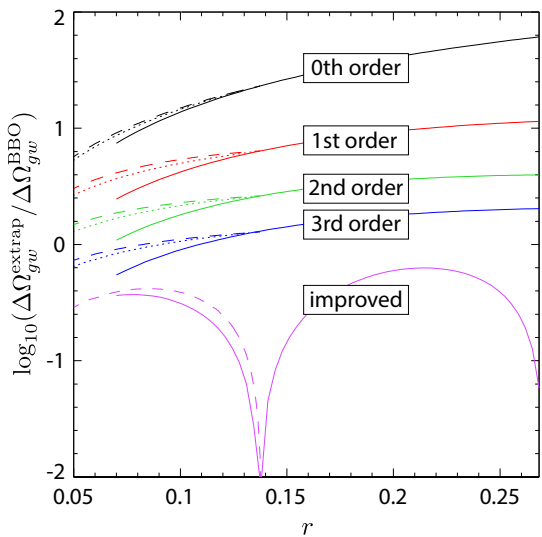}}
\caption{To quantify the accuracy of the extrapolations discussed in 
the text, we compare $\Delta\Omega_{gw}^{{\rm BBO}}=10^{-17}$, the 1-sigma error 
bar of ``standard BBO" after one year of operation, with $\Delta\Omega_{gw}^{{\rm extrap}}
=|\Omega_{gw}^{{\rm exact}}-\Omega_{gw}^{{\rm extrap}}|$, the difference between 
the exact value $\Omega_{gw}^{{\rm exact}}$ on BBO scales (predicted by a
given potential) and the extrapolated value $\Omega_{gw}^{{\rm extrap}}$
based on the the zeroth-order (black), first-order (red), second-order (green)
or third-order (blue) traditional extrapolation (\ref{lnkExpansion}), or the 
improved extrapolation (\ref{sqrtV}) (pink).  We compare with three families of 
standard inflationary potentials: (i) the monomial potentials $V(\varphi)=\lambda\varphi^{n}$
with $1<n<4$ (solid curves); (ii) the axion-like (``natural inflation") potentials
$V(\varphi)=V_{0}[1-{\rm cos}(\varphi/f)]$ (dashed curves); and (iii) the higgs-like 
potentials $V(\varphi)=V_{0}[1-(\varphi/f)^{2}]^{2}$ (dotted curves).  The zeros in 
the pink solid curve at $r\approx0.14$ and $r\approx0.27$ correspond to the potentials
$V=(1/2)m^{2}\varphi^{2}$ and $V=\lambda\varphi^{4}$, where our improved extrapolation
is essentially exact.  The axion curve dips to zero at $r\approx0.14$, too: this happens because,
in the limit $f\to\infty$, the final 60 e-folds of inflation occur near the bottom of the 
potential well, where the potential may be well appoximated by $(1/2)m^{2}\varphi^{2}$.  
Finally, the improved extrapolation is also nearly exact for the higgs-like potential (iii), so the 
pink dotted curve lies below the bottom of the figure, and is not shown.}
\label{fig:Extrap}
\end{figure}

To quantify the accuracy of the traditional extrapolation, in Fig.~\ref{fig:Extrap} we show how 
well the exact predictions on BBO scales from three standard families of inflationary potential
are approximated by including terms up to 0th, 1st, 2nd, or 3rd order in Eq.~(\ref{lnkExpansion}).
We see that this series converges rather slowly (each additional 
order only improves the accuracy by a factor of $\sim2$) and it gives a relatively
inaccurate prediction on BBO scales (in the sense that the 2nd order extrapolation, which is
the highest order that can be realistically used in the absence of a measurement of the scalar tilt $\alpha_{s}$, gives an extrapolation error which dominates over the the BBO one-year instrumental error).  The reason that the traditional extrapolation behaves so poorly, even when applied to the simplest inflationary potentials, is that although we are trying to extrapolate from CMB scales (which left the horizon $\sim60$ e-folds before the end of inflation) to LI scales which left the horizon $\sim40$ e-folds later, the traditional extrapolation (\ref{lnkExpansion}) only uses information from CMB scales, while ignoring two key facts about the end of inflation.  (i) First, since the horizon wavenumber $k=aH$ varies non-monotonically (it grows during inflation, 
reaches its maximum at the end of inflation, and then shrinks after inflation), the derivative
$P_{t}'(k)$ must diverge to $-\infty$ at the end of inflation, a feature that a finite Taylor series in ${\rm ln}\,k$ can 
never capture; we can resolve this problem by expanding in $\varphi$ rather 
than ${\rm ln} k$, since $\varphi$ varies monotonically through the end of 
inflation.  (ii) Second, after the end of inflation, $\varphi$ settles to a 
minimum with $V\approx 0$; we can incorporate this behavior into our approximation by  
expanding $V^{1/2}$ in $\varphi$. This ensures that $V(\varphi)$ is a non-negative function with a stable minimum and, moreover, for $r<-(16/3)\delta n_{s}$ (as is already suggested by the observations), 
$V^{1/2}(\varphi)$ crosses through zero [see Eq.~(\ref{sqrtV})] which implies that the minimum of $V(\varphi)$ is automatically at $V=0$, as desired.  Thus, our  proposal is to expand $V^{1/2}$ as a series in $\varphi$, rather than expanding ${\rm ln}\,P_{t}$ as a series in ${\rm ln}\,k$. As we shall see, by accomodating the end of inflation in this way, we obtain a much more accurate extrapolation from CMB to LI scales.

\begin{figure}[t]
\centerline{\includegraphics[width=7.5cm,clip=True,trim=0cm 5cm 0cm 5cm]{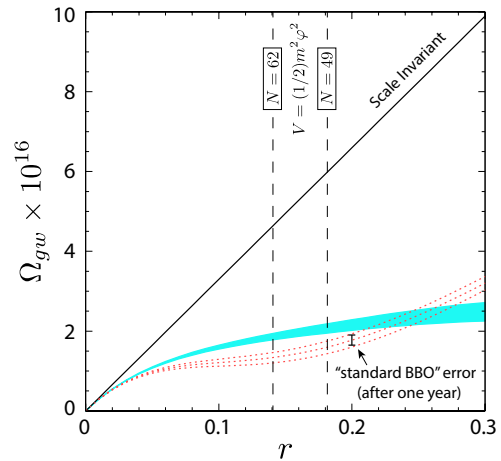}}
\caption{$\Omega_{gw}$ (on BBO scales) vs $r$ (on CMB scales).  
The solid black diagonal line shows the relationship if $P_{t}$ is 
exactly scale invariant ($n_{t}=0$).  The filled cyan band shows the third-order 
traditional extrapolation (\ref{lnkExpansion}), for Planck's best fit value 
$n_{s}=0.9603$, and $\alpha_{s}=0\pm0.0003$ (this range is chosen to make the 
width of the cyan band comparable to the size of the BBO error bar).  The orange 
dotted curves show the improved extrapolation based on (\ref{sqrtV}), the same $n_{s}$, 
and a future $1\sigma$ error bar ($n_{s}=0.9603\pm0.0017$).  The vertical dashed lines 
show the range of $r$ predicted by $V=(1/2)m^{2}\varphi^{2}$ inflation, for $49<N<62$,
{\it cf.}\ \cite{Creminelli:2014oaa}.  For reference, we also show the $1\sigma$ error bar for 
``standard BBO" after one year of operation.}
\label{fig:rVsOmegaGW}
\end{figure}

Just as we used the CMB observables $\{P_{s}, r, n_{s}\}$ to determine 
${\rm ln}\,P_{t}$ up to quadratic order in ${\rm ln}(k)$, we can use them to 
determine $V^{1/2}$  up to quadratic order in $\varphi$.  Specifically, if (without
loss of generality) we take $\phi=0$ to be the field value corresponding to 
$k_{0}$ and define $V_{0}\equiv V(\varphi_{0})$, 
$\epsilon_{V}^{}\equiv(1/2)M_{pl}^{2}(V_{0}'/V_{0})^{2}$, and 
$\eta_{V}^{}\equiv M_{pl}^{2}(V_{0}''/V_{0})$, then we can write
\begin{equation}
  \label{sqrtV}
  V^{1/2}(\varphi)=V_{0}^{1/2}\left[1-\sqrt{\frac{\epsilon_{V}}{2}}\frac{\varphi}{M_{pl}}
  +\frac{\eta_{V}-\epsilon_{V}}{4}\frac{\varphi^{2}}{M_{pl}^{2}}\right].
\end{equation}
Here $V_{0}$, $\epsilon_{V}$ and $\eta_{V}$ are given, to NLO in slow roll, by
\begin{subequations}
  \begin{eqnarray}
    \frac{V_{0}}{M_{pl}^{4}}\!&\!=\!&\!12\pi^{2}P_{s}\tilde{r}[1+(C+\frac{5}{6})\tilde{r}], \\
    \epsilon_{V}\!&\!=\!&\!\frac{1}{2}\tilde{r}[1+(C-\frac{1}{3})(\tilde{r}+\delta n_{s})], \\
    \eta_{V}\!&\!=\!&\!\frac{\delta n_{s}}{2}\!+\!\frac{3\tilde{r}}{2}\!-\!\frac{7}{12}\tilde{r}^{2}
  \!+\!\Big(\frac{C}{4}\!-\!1\Big)\tilde{r}\delta n_{s}\!-\!\frac{1}{12}\delta n_{s}^{2},
  \end{eqnarray}
\end{subequations}
where $C\!\equiv\!-2\!+\!{\rm ln}(2)\!+\!\gamma$ and $\gamma\!\approx\!0.577216$ is Euler's constant.  (For help in deriving these expressions, see Refs.~\cite{Lidsey:1995np, Liddle:1994dx}.)  Then we can solve the background equations $\ddot{\varphi}+3H\dot{\varphi}+V'(\varphi)=0$ and $3M_{pl}^{2}H^{2}=\frac{1}{2}\dot{\varphi}^{2}+V(\varphi)$ exactly ({\it i.e.}\ numerically) for this potential, and compute the corresponding tensor spectrum at NLO in slow roll \cite{Lidsey:1995np, Cortes:2006ap}, $P_{t}=8[1-(C+1)\epsilon_{H}^{}]^{2}(H/2\pi M_{pl})^{2}$, where the Hubble slow roll parameter $\epsilon_{H}\equiv 2 M_{pl}^{2}[H'(\varphi)/H(\varphi)]^{2}$.  

In Fig.~\ref{fig:Extrap}, we show how the extrapolation method based on Eq.~(\ref{sqrtV}) provides a much-improved appoximation to the exact predictions for our three standard families of inflationary potentials.  We see that this improved extrapolation beats the traditional 2nd order extrapolation by roughly an order of magnitude, and even beats the traditional 3rd order extrapolation by a factor of several.  In particular, the improved extrapolation error is smaller than the BBO instrumental error (unlike the traditional extrapolation error which, as discussed above, is larger than the BBO instrumental error, even for the simplest and smoothest inflationary potentials). 

More generally, we expect the extrapolation based on Eq.~(\ref{sqrtV}) to outperform the traditional extrapolation for the broader class of inflationary potentials $V(\varphi)$ having the two key features that one usually requires in a successful model of single field inflation: namely, (i) that inflation ends by the usual single-field mechanism in which the potential steepens until $\epsilon_{H}=1$, and (ii) that after inflation, the field comes to rest at a potential minimum with $V=0$ ({\it i.e.}\ with  vanishing, or nearly vanishing, residual cosmological constant after the end of inflation).  The reason is simply that the improved extrapolation incorporates more information: whereas both the traditional and improved extrapolations take into account the same information about the shape of the inflaton potential on CMB scales (inferred from measurements of $P_{s}$, $r$ and $n_{s}$), the traditional extrapolation ignores (and inevitably violates) the two end-of-inflation boundary conditions discussed above, while the improved extrapolation automatically incorporates them.  On the other hand, it is important to emphasize that, if one is free to choose an arbitrary function $V(\varphi)$, then it is still always possible to ``break" any extrapolation scheme, {\it e.g.}\ by inserting a transient feature in $V(\varphi)$ between CMB and BBO scales.  (In particular, such a transient would break both the traditional and the improved extrapolations schemes.)  The point, then, is not that the improved extrapolation based on Eq.~(\ref{sqrtV}) is unbreakable; but, rather, that insofar as the traditional extrapolation is a reasonable one, the extrapolation based on Eq.~(\ref{sqrtV}) may be expected to perform even better.

Fig.~\ref{fig:rVsOmegaGW} shows the value of $\Omega_{gw}$ on BBO scales,
as predicted by the various extrapolations we have discussed.
Note, in particular, that BBO should be able to distinguish with very high 
significance between the inflationary prediction (the dotted orange curves) and
the prediction based on a strictly scale invariant tensor spectrum with $n_{t}=0$ 
(the solid black line).  Thus, the CMB and BBO in combination can provide a much more 
stringent test of the inflationary consistency relation than the CMB alone. 
Also note that, due to the large separation between $k_{{\rm CMB}}$ and 
$k_{{\rm BBO}}$, even a small decrease of $\alpha_{s}$ on CMB scales can 
lead to a detectable suppression of $\Omega_{gw}$ on BBO scales, as shown 
by the cyan band.  This sensitivity of $\Omega_{gw}$ to $\alpha_{s}$ should allow 
the combination of CMB and BBO measurements to place very stringent 
constraints on $\alpha_{s}$ \cite{Smith:2014kka}.  

One might hope that BBO could be used to distinguish between different inflationary potentials that make indistinguishable predictions on CMB scales. In practice, though, when two simple and smooth potentials (without sharp features) make indistinguishable predictions on CMB scales, we find that they also tend to make nearly indistinguishable predictions on BBO scales.\footnote{A number of previous papers ({\it e.g.}\ \cite{Smith:2006xf,Chongchitnan:2006pe,Caligiuri:2014ola}) have used Monte-Carlo/Hubble-flow techniques to investigate how constraints on CMB scales relate to constraints on laser interferometer scales; and, at first glance, the figures in those papers may appear to suggest that extrapolating from CMB scales to laser interferometer scales leads to wider range of predictions for the direct detection amplitudes compared with the extrapolation uncertainty that we seem to find.  But one must be careful to compare apples to apples!  First of all, it is important to note that when those papers select a set of relevant inflationary trajectories $H(\varphi)$, they do not impose our second condition (ii), and two of them \cite{Smith:2006xf,Chongchitnan:2006pe} also do not impose our first condition (i); if they did impose these extra conditions, it would winnow down their set of relevant trajectories $H(\varphi)$ greatly, and their corresponding predictions would be correspondingly tightened.  Second, those papers consider the spread in the direct detection prediction, given a particular (sometimes quite large) spread in the CMB constraints; whereas we have been looking at the error in the direct detection prediction, given (perfectly) known values for the CMB observables (since we want to clearly disentangle this uncertainty from the one coming from propagation of the observational uncertainty on CMB scales).}  For example, 
consider two standard potentials: the axion-like potential $V(\varphi)=V_{0}[1-{\rm cos}(\varphi/f)]$ 
and the higgs-like potential $V(\varphi)=V_{0}[1-(\varphi/f)^{2}]^{2}$.  For either of 
these two potentials, as we vary the three parameters $V_{0}$, $f$, and $49<N_{{\rm cmb}}<62$
(subject to the constraint imposed by the observed amplitude of the scalar power 
spectrum), the corresponding predictions for $r$ and $n_{s}$ fill out a 2-dimensional 
swath in the $\{n_{s},r\}$ plane. (For example, the axion-like swath is shown in purple
in Fig.~1 of Ref.~\cite{Ade:2013uln}.)  Since these two swaths overlap, we can ask: 
when the parameters are chosen so that the axion and higgs potentials make exactly 
the same prediction for $(n_{s},r)$, do they make distinguishable predictions on
BBO scales?  The answer is no: $\Omega_{gw}^{{\rm higgs}}$ and $\Omega_{gw}^{{\rm axion}}$ 
always differ by less than $\Delta\Omega_{gw}^{{\rm BBO}}$, the 1$\sigma$ error bar for ``standard BBO." 

{\bf Step (ii): The tensor transfer function:}
In Figs.~\ref{fig:Extrap} and \ref{fig:rVsOmegaGW}, to propagate 
the primordial power spectrum $P_{t}(k_{{\rm BBO}})$ forward to the present time,
we assumed that the reheat temperature ({\it i.e.} the temperature at the start of the 
ordinary radiation-dominated era) was higher than $10^{4}{\rm TeV}$, which is then 
the temperature at which BBO's gravitational waves re-entered the horizon in the early 
universe.  We further assumed that ever since that time ({\it i.e.}\ for temperatures 
$T\!\lesssim\!10^{4}{\rm TeV}$), the effective numbers of relativistic species 
$g_{\ast}(T)$ and $g_{\ast,s}(T)$ \cite{Kolb:1990vq} were given by their standard model values,
so that we could use the standard tensor transfer
function on these scales \cite{Smith:2005mm, Boyle:2005se, Boyle:2007zx}.  It is important to note, 
though, that the tensor transfer function could deviate from this standard expectation for a variety 
of interesting reasons; and that observing such deviations with an experiment like BBO could 
provide qualitatively new clues about ultra-high energy scales and early times that 
are not probed in any other way.   For example, BBO would be sensitive to the density 
of relativistic free-streaming particles at $T\!\sim\!10^{4}{\rm TeV}$: see \cite{Boyle:2005se}
for details and other examples.


BICEP2's claimed detection of primordial gravity waves is an opportunity to
reconsider the science case for BBO and DECIGO.  In this paper, we have 
emphasized and quantified how BBO+CMB can test the predictions
of inflation to a far greater extent than the CMB alone.  It is also 
important to emphasize that, 5 years after BBO was proposed as a 
mission to detect the inflationary gravitational wave background, it 
was realized that it also had a very remarkable and compelling 
science case that had nothing to do with inflation \cite{Cutler:2009qv, Hirata:2010ba}.
Together, these reinforce the case for BBO or something like it as an 
extremely well motivated project that deserves wider and more 
serious attention than it has received to date.

\vskip 0.2cm

{\em Acknowledgements.}
We wish to thank Mark Halpern and Gary Hinshaw for discussions. 
Research at Perimeter Institute is supported by the Government of Canada
through Industry Canada and by the Province of Ontario through the Ministry of Research \& Innovation.
LB and KMS were supported by NSERC Discovery Grants.
CD was supported by the National Science Foundation grant number AST-0807444, NSF grant number PHY-088855425, and the Raymond and Beverly Sackler Funds.

\bibliographystyle{h-physrev}
\bibliography{r02consistency}

\end{document}